\documentclass{ws-ijmpa}

\input BoxedEPS
\SetRokickiEPSFSpecial  
\HideDisplacementBoxes

\newcommand{\fract}[2]{{\textstyle \frac{#1}{#2}}}

\begin{document}

\newcommand{\be}{\begin{equation}}
\newcommand{\ee}{\end{equation}}
\newcommand{\bea}{\begin{eqnarray}}
\newcommand{\eea}{\end{eqnarray}}
\newcommand{\ben}{\begin{enumerate}}
\newcommand{\een}{\end{enumerate}}
\newcommand{\bit}{\begin{itemize}}
\newcommand{\eit}{\end{itemize}}
\newcommand{\la}[1]{\label{#1}}
\newcommand{\half}{\frac{1}{2}}
\newcommand{\dk}{\frac{d^{4}k}{(2\pi)^4i}}
\newcommand{\di}[1]{\frac{d^{4}{#1}}{(2\pi)^4i}}
\newcommand{\eq}[1]{eq.~(\ref{#1})}
\newcommand{\Eq}[1]{Eq.~(\ref{#1})}
\newcommand{\eqs}[2]{eqs.~(\ref{#1}) and (\ref{#2})}
\newcommand{\f}[2]{\tilde {F}_{#1}(#2)}
\newcommand{\ns}{\hspace{-0.5ex}}
\newcommand{\p}{p \ns \cdot \ns \gamma }
\newcommand{\hf}{{\textstyle \frac{1}{2}}}
\newcommand{\Tr}{\,\mbox{Tr}\,}
\newcommand{\trc}{\,{\mbox{Tr}}_C\,}
\newcommand{\tr}{\,\mbox{tr}\,}
\newcommand{\Ln}{\,\mbox{Ln}\,}
\renewcommand{\ln}{\,\mbox{ln}\,}
\renewcommand\Im{{\rm Im}}
\newcommand\mn{{\mu\nu}}
\newcommand\parm{\par\medskip}
\renewcommand\Re{{\rm Re}}
\newcommand{\cl}{\centerline}

\def\pmb#1{\setbox0=\hbox{$#1$}%
\kern-.025em\copy0\kern-\wd0
\kern.05em\copy0\kern-\wd0
\kern-.025em\raise.0433em\box0 }

%

\def\nocropmarks{\vskip5pt\phantom{cropmarks}}

\let\trimmarks\nocropmarks      

%

\markboth{R. L. Jaffe}
{Open Questions in High Energy Spin Physics}

%
\catchline{}{}{}
%

\setcounter{page}{1}

\title{OPEN QUESTIONS IN HIGH ENERGY SPIN PHYSICS\\}

\author{\footnotesize ROBERT L. JAFFE}

\address{Center for Theoretical Physics and Department of Physics \\
Laboratory for Nuclear Physics, 
Massachusetts Institute of Technology\\
Cambridge, Massachusetts 02139\footnote{Email:  jaffe@mit.edu\quad MIT-CTP-3230\quad
hep-ph/0201068} }

\maketitle


\begin{abstract}
I describe a few of the most exciting open questions in high
energy spin physics.  After a brief look at $(g-2)_\mu$ and the
muon electric dipole moment, I concentrate on QCD spin physics.
Pressing questions include the interpretation of new asymmetries
seen in semi-inclusive DIS, measuring the polarized gluon and
quark transversity distributions in the nucleon, testing the
DHGHY Sum Rule,  measuring the orbital angular momentum in the
nucleon, and many others which go beyond the space and time
allotted for this talk.
\end{abstract}

\section{Introduction}
I would like to thank the organizers for the opportunity to deliver
the opening talk at this exciting conference.  When I was last in
Beijing in 1981 no one could have predicted what lay just around the
corner in the domain of high energy spin physics.  The measurement of
the quark spin contribution to the nucleon spin by the European Muon
Collaboration in the mid-1980s launched a new era in QCD spin
physics, which will be the principal focus of my talk.

The organizers asked me to stress open questions.  This has its
advantages, since they did not require me to provide answers.  Recent
excitement leads me to mention very briefly some headlines in physics
beyond the Standard Model.  After that I will get down to the business
of QCD\null. Of necessity, I have singled out a few topics for attention. Other issues of equal, perhaps some would say greater, interest will
only be mentioned in passing along with some areas which I have found
particularly frustrating.  Here is my outline:

\begin{romanlist}
	\item Headlines:  $g-2$ and the muon electric dipole moment
	\item Focus: High energy spin physics in QCD
		\begin{romanlist}
			\item What are the origins and implications of the Hermes azimuthal 			asymmetry?   			\item What is the quark transversity distribution in the nucleon?
			\item What is the polarized gluon distribution in the nucleon?
			\item Can the Gerasimov-Drell-Hearn-Hosata-Yamamoto Sum Rule fail?
			\item What can be said about quark and gluon orbital angular 			momentum in the nucleon?
		\end{romanlist}
	\item In passing\ldots (revisited briefly at the end)
		\begin{romanlist}			
			\item What happens to $g_{1}(x,Q^{2})$ at very low $x$?\footnote{
			This topic was discussed in detail but is only mentioned briefly in the 			proceedings due to space limitations.}
			\item What are the properties of the $s$ and $\bar s$ quarks in the 			nucleon?$^{\rm a}$
			\item What are the spin-dependent quark-gluon correlations in the 			nucleon? 			\item What is the flavor decomposition of the nucleon's 			spin?\footnote{This topic is omitted entirely from
			the proceedings due to space limitations.}
			\item How do QCD spin effects behave as $Q^{2}\to 0$?$^{\rm b}$
			\item What can ``off-forward'' parton distributions tell us about 			hadron properties?
			\item Can some deeper order be found in the proliferation of 	
$\vec k_{\perp}$ and $x$ dependent distribution and fragmentation 		
functions?
		\end{romanlist}
\end{romanlist}

\section{Headlines:  $g-2$ and the muon electric dipole moment}

No discussion of high energy spin physics would be complete without
recognizing the important role of spin in precision tests of the
Standard Model.  Recent headlines include a new high precision
measurement of $(g-2)_{\mu}=2a_{\mu}$ and a proposal to increase the
precision on the muon's EDM by several orders of magnitude.

The muon's magnetic moment probes certain extensions of the Standard
Model up to energies equivalent to LEP and the Tevatron, and is
sensitive to SUSY and other novelties.  After years of hard work and
great patience, the Brookhaven experiment (E821) has reported a value
for $a_{\mu}$ which will challenge the Standard Model.  It is
conventional to quote values for $a_{\mu}$ in units of $10^{-10}$ or
$10^{-11}$ and accuracy in parts per million.  Thus the old CERN
$\mu^{+}$ value of $a_{\mu}\times 10^{10}=116\ 591\ 00(110)$ has an
accuracy of 10 ppm.\cite{Bailey:1979mn} The number reported from BNL,
$a_{\mu}^{\rm BNL}\times 10^{10}=116\ 592\ 02(14)(6)$, has an accuracy
of 1.3 ppm.\cite{Brown:2000sj} The new weighted average of data on
$a_{\mu}$ disagrees with the ``standard'' theoretical estimate,
$a_{\mu}^{\rm TH}\times 10^{11}=116\ 591\ 596(67)$ by 2.6 standard
deviations.  At present the precision of the theoretical estimate of
$(g_{\mu}-2)$ is principally limited by the lack of information on
higher order QCD contributions, which require further
study.\cite{Melnikov:2001uw} Approved experiments at BNL plan to
reduce the statistical and systematic uncertainties on $a_{\mu}$ to
about 0.3 ppm, making better understanding of the QCD contribution a
very high priority.\footnote{Between the conference and the
preparation of the proceedings, new estimates of the QCD
contributions, and in particular, a sign reversal in the contribution
of hadronic light-by-light scattering, have reduced the discrepancy
between theory and experiment to 1.6$\sigma$.\cite{Knecht:2001qf,Knecht:2001qg,Blokland:2001pb} Improved
understanding of QCD contributions could increase the impact of
further $(g-2)_{\mu}$ measurements by decreasing the uncertainty in
the Standard Model prediction.}

Electric dipole moments (EDMs) probe CP violation, one of the most
poorly understood aspects of the Standard Model.  If all CP-violation
is encoded in the CKM matrix, EDMs are too small to measure. Therefore EDMs are an excellent place to look for CP-violation beyond
the Standard Model.\cite{Khriplovich:2000zh}  Attempts to explain the
origin of the baryon excess in the Universe suggest that other sources of
CP-violation may be waiting to be discovered.

Standard Model (i.e., CKM) predictions for EDMs are far smaller than
the reasonable goals of experiments.  
This need not be the case for interesting alternatives.  In fact one might
expect a new physics contribution to the anomalous magnetic moment,
$a_{\mu}$ and the EDM, $d_{\mu}$, to be comparable.\cite{} 
 More precisely, $a_{\mu}$ and $d_{\mu}$ are defined by
\begin{equation}
	{\cal L}_{\rm new} = \Bigl(\frac{e}{4m_{\mu}}\Bigr)a_{\mu}\bar\mu
	\sigma^{\alpha\beta} F_{\alpha\beta}\mu
	-\frac{i}{2}d_{\mu}\,\bar\mu
\sigma^{\alpha\beta}F_{\alpha\beta}\gamma_{5}\mu 
\end{equation}
so one might expect $a_{\mu}$ and $2m_{\mu}\,d_{\mu}/e$ to be in
proportion to $\tan\phi_{\rm CP}$, where $\phi_{\rm CP}$ is a new CP
violating phase.\cite{Marciano,Graesser:2001ec,Feng:2001sq} So EDM
measurements at a sensitivity comparable to existing limits on
$a_{\mu}$ could provide fertile ground in which to look for new
sources of CP-violation.  Of particular interest is Semertzidis's
proposal to use the BNL $(g-2)$ ring to improve the limit on the muon
EDM by as much as six orders of magnitude\cite{Semertzidis:1998sp}
beyond the present limit of order $10^{-18}$ e-cm.  Taking $\phi_{\rm
CP}\sim 1$, $d_{\mu}\sim 10^{-22}$ e-cm, so this is an interesting
possibility.  Readers interested in this simple and elegant idea
should consult the review by Khriplovich.

\section{Focus:  High Energy Spin Physics in QCD}

\subsection{Origins Implications of the Hermes Azimuthal Asymmetry?}

To my mind the single most interesting development in QCD spin physics
reported in the past two years is the azimuthal asymmetry in pion
electroproduction from Hermes.\cite{Airapetian:2000tv} It is
interesting in itself and also as an emblem of a new class of spin
measurements involving spin-dependent fragmentation processes, which
act as filters for exotic parton distribution functions like
transversity.

Fragmentation functions allow us to access and explore the spin
structure of unstable hadrons, which cannot be used as targets for
deep inelastic scattering.  Examples include the longitudinal and
transverse spin dependent fragmentation functions of the $\Lambda$,
schematically $\vec q_{\parallel}\to \vec \Lambda_{\parallel}$ and
$\vec q_{\perp}\to\vec \Lambda_{\perp}$.  Since the $\Lambda\to p \pi$
decay is self-analyzing it is relatively easy to measure the spin of
the $\Lambda$.  By selecting $\Lambda$'s produced in the current
fragmentation region one can hope to isolate the fragmentation process
$q\to\Lambda$.  Another, perhaps less obvious, example is the tensor
fragmentation function of the $\rho$, denoted schematically by
$(q\to\rho_{\pm})-(q \to\rho_{0})$, where $\rho_{h}$ are $\rho$ helicity
states.\cite{Schafer:1999am} $\rho$ decay transmits no spin
information, but it distinguishes the longitudinal and transverse
helicity states required for this measurement.  Such data are already
available.  The challenge to theorists is to make use of it.

\begin{figure}[ht]
\centerline{\BoxedEPSF{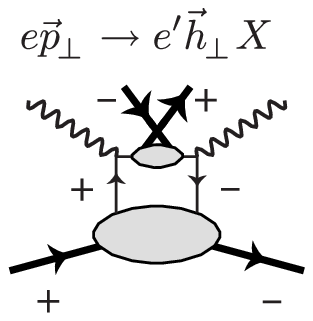 scaled 1000}}
\caption{}
\label{decouple}
\vspace*{-\bigskipamount}
\end{figure} 

Even if we do not know how to interpret fragmentation functions, we
can use them as filters, to select parton distribution functions which
decouple from completely inclusive DIS. The salient example is the use
of a helicity flip fragmentation function to select the quark
transversity distribution.  As shown in Fig.~\ref{decouple}, by
interposing a helicity flip fragmentation function on the struck quark
line in DIS, it is possible to access the transversity (see below). There are several candidates for the necessary helicity flip
fragmentation function:

\begin{romanlist}
	
	\item $e \vec p_{\perp}\to e' \vec\Lambda_{\perp} X$ 	
	In this case the helicity flip fragmentation function of the
	$\Lambda$ is exactly analogous to the transversity distribution
	function in the nucleon.\cite{Kunne:1993nq,Jaffe:1996wp} The only
	difficulty with this example is the relative rarity of $\Lambda$'s
	in the current fragmentation region, and the possibly weak
	correlation between the $\Lambda$ polarization and the
	polarization of the $u$ quarks which dominate the proton.
	
	\item $e \vec p_{\perp}\to e' \pi(\vec
	k_{\perp})X$\cite{Collins:1994kq} [The ``Collins Effect'']
		
	In this case the azimuthal angular distribution of the pion
	relative to the $\vec q$ axis can be analyzed to select the
	interference between pion orbital angular momentum zero and one
	states. This observable correlates with quark helicity flip.  In more traditional
	terms the effect is proportional $\vec S_{\perp}\cdot \vec
	q\times\vec p_{\pi}$.  This is multiplied by the quark
	transversity in the target and an unknown fragmentation function
	(known as the Collins function) describing the propensity of the
	quark to fragment into a pion in a superposition of orbital
	angular momentum zero and one states.  The fact that fragmentation
	functions depend on $z$ while distribution functions depend on $x$
	allows the shape of the transversity distribution to be measured
	in this manner.
	
	\item $e \vec p_{\perp}\to e' \pi\pi X$\cite{Collins:1994ax,Jaffe:1998hf}
	
	In this case the angular distribution of the two pion final state
	substitutes for the azimuthal asymmetry.

	\end{romanlist}
\vspace*{-\smallskipamount}

Last year Hermes announced the observation of an azimuthal asymmetry
similar to the Collins asymmetry described above, but with a
longitudinally polarized target: $e \vec p_{\parallel}\to e' \pi(\vec
k_{\perp})X$.  Their data are shown in T.~Shibata's contribution to
this conference.  This asymmetry could be a (suppressed) reflection of
the Collins effect because the target spin, while parallel to the
electron beam, has a small component, ${\cal O}(1/Q)$ perpendicular to
the virtual photon.  It could also result from competing twist-three
helicity flip effects also suppressed by $1/Q$.  Unless the Hermes
asymmetry is entirely twist-three, which seems unlikely, it appears
that the prospects for observing a large azimuthal asymmetry from a
{\it transversely\/} polarized target are very good.  Hermes will be
running with a transversely polarized target this year and their
results will be awaited with considerable excitement.  COMPASS has similar objectives.  The spin program at RHIC hopes to access transversity by observing a similar $(\pi\pi)$ azimuthal asymmetry in $pp$ collisions.
 \subsection{The Quark Transversity Distribution in the Nucleon?}

One of the major accomplishments of the recent renaissance in QCD spin
physics has been the rediscovery and exploration of the quark {\it
transversity distribution}.  First mentioned by Ralston and Soper in
1979 in their treatment of Drell-Yan $\mu$-pair production by
transversely polarized protons,\cite{RS1979} the transversity was not
recognized as a major component in the description of the nucleon's
spin until the early 1990s.\cite{artru90,jaffe91,cortes92,jaffe95} We now
know that the transversity, $\delta q(x,Q^{2})$, together with the
unpolarized distribution, $q(x,Q^{2})$, and the helicity distribution,
$\Delta q(x,Q^{2})$, are required to give a complete description of
the quark spin in the nucleon at leading twist.  One equation tells
this story clearly:
\begin{equation}
{\cal A}(x,Q^2)\! =\! \fract12  q(x,Q^2)\,I\otimes I + \fract12  \Delta
q(x,Q^2)\,\sigma_3 \otimes \sigma_3 +\fract12   \delta q(x,Q^2) 
(\sigma_+\otimes\sigma_-+\sigma_-\otimes\sigma_+)\, 
\label{symmetry}
\end{equation}
Here, ${\cal A}$ is the quark distribution in a nucleon as a density
matrix in both the quark and nucleon helicities (hence the direct
product of two Pauli matrices in each term).  $q$ governs spin average
physics, $\Delta q$ governs helicity dependence, and $\delta q$
governs helicity flip -- or transverse polarization -- physics.

The transversity can be interpreted in parton language as the
probability to find quarks of momentum fraction $x$, transversely
polarized in a transversely polarized nucleon at infinite momentum. The quark momentum distribution is well known and the helicity
distribution is becoming better known.  In contrast nothing is known
about transversity from experiment, because it decouples from
inclusive DIS on account of a selection rule.  At leading twist
helicity and chirality are identical.  Transversity corresponds to
helicity, and therefore chirality, flip.  So transversity decouples
from processes with only vector or axial vector couplings.  In order
to access transversity it is necessary to flip a quark's helicity and
then flip it back in two soft processes.  Two examples where
transversity does not decouple are transversely polarized Drell-Yan: $\vec
p_{\perp} \vec p_{\perp}\to \mu^{+}\mu^{-} X$ (the original
Ralston-Soper process where transversity was discovered) and
semi-inclusive DIS where a final state fragmentation function flips
helicity, $e\vec p_{\perp}\to e'\vec h_{\perp}X$, as shown in
Fig.~\ref{decouple}.  Measurements of quark transversity rank
high on the agendas of Hermes, COMPASS and RHIC.

\subsection{The Polarized Gluon Distribution in the Nucleon?}

A {\it direct\/} measurement of the polarized gluon distribution in
the nucleon is probably the highest priority for QCD spin physics. The first, {\it indirect\/} estimates of $\Delta G(x,Q^{2})$ have been
made by the SMC group by studying the $Q^{2}$ dependence of the quark
distribution, $\Delta q(x,Q^{2})$, which couples to $\Delta G$ through
renormalization group evolution.  [See Ref.~\cite{Adeva:1998vw} for
details of the process and references to the original literature.] Further refinement of the indirect method will improve our knowledge
of $\Delta G$ , but direct measurement is essential to determine gross
features of its shape.

Several direct methods are being pursued: 
\begin{romanlist}
	\item $\bar c c$ pair production in $e\vec p_{\parallel}\to
	e'(\bar c c)X$ and related methods.
	
	The COMPASS Collaboration has proposed to extend this powerful
	probe of the unpolarized gluon distribution to the polarized
	case.\cite{Bradamante:2000yc} The basic mechanism is photon-gluon
	fusion.  Variations on this method include two jet production:
	$e\vec p_{\parallel}\to e' \hbox{ jet jet } X$ at large transverse
	momentum (as originally envisioned by Carlitz, Collins, and
	Mueller\cite{Carlitz:1988ab}); $\bar c c$ photoproduction $\gamma
	\vec p_{\parallel}\to (\bar c c)X$; and pion pair production,
	$\gamma\vec p_{\parallel}\to \pi\pi X$, which Hermes hopes to use
	a lower center of mass energies where $\bar c c$ and two jet
	production are not available.\cite{Vincter:2000pc}
	
	\item Single photon production at high transverse momentum in
	polarized $\vec p_{\parallel} \vec p_{\parallel}\to \gamma \hbox{
	jet } X$ and related methods.
	
	This is a prime goal for the polarized proton program at
	RHIC.\cite{Bunce:2000uv} Here the basic mechanism is the QCD
	Compton process.  This process should be an excellent probe of the
	polarized gluon distribution.  However there is some controversy
	about higher order QCD corrections which has yet to be resolved in
	the unpolarized case.  Variations replace the high energy photon
	with a jet, or in the case of poor jet acceptance, a leading pion
	at high transverse momentum.
\end{romanlist}

Estimates of the precision of these methods have become available as
better simulations come on line for COMPASS and RHIC. For detailed projections see the talks by T.~Morii, N.~Saito, and T.~Shibata in these proceedings. 

\subsection{The Gerasimov-Drell-Hearn-Hosada-Yamamoto  Sum
Rule?}

The prospects for a definitive test of this deep and ancient sum
rule\cite{Gerasimov:1966et,Drell:1966jv,Hosoda:1966} are now
excellent.  Old studies of resonance contributions to the sum rule
indicate that the sum rule is approximately saturated, but data in the
Regge region are crucial.  Experiments proposed and/or underway in
Bonn (at ELSA), Mainz (at MAMI) and at JLab will cover a wide range of
energies with high polarization and high statistics.
 The question I would like to raise here is ``What does
the DHGHY Sum Rule test?''.  The sum rule reads 
\be
	\frac{2\pi^{2}\alpha}{M ^{2}}\kappa^{2} = \int_{0}^{\infty} 	
\frac{d\nu}{\nu}\bigl(\sigma_{P}(\nu)-\sigma_{A}(\nu)\bigr)
	\la{DHGHYI}
\ee
where $\kappa$ and $M$ are the anomalous magnetic moment and mass of
the target, and $\sigma_{P,A}$ are the total photoabsorption cross
sections (as functions of the laboratory photon energy, $\nu$) for
target and photon spins parallel and antiparallel.

The sum rule rests on two assumptions: first Low's low energy theorem
$f_{2}(0)=-\fract12 \frac{\alpha}{M^{2}}\kappa^{2}$, where $f_{2}(0)$ is
the energy derivative of nucleon's forward spin-flip Compton amplitude
at zero energy;\cite{Low:1954kd} and second, the assumption that $f_{2}(\nu)$
obeys an unsubtracted dispersion relation.  Low's theorem relies only
on gauge invariance and analyticity, and is not expected to be
violated.  The dispersion relation reads 
\be
	\hbox{Re}\,f_{2}(\nu) =\sum_{j=0}^{J_{\rm MAX}}c_{j}\nu^{2j}+
	\frac{1}{8\pi^{2}}\hbox{P}\int_{0}^{\infty}
	d\nu'^{2}\,\frac{\sigma_{A}(\nu')-\sigma_{P}(\nu')}
	{\nu'^{2}-\nu^{2}} \ .
	\la{disp}
\ee
The polynomial is usually omitted in writing the dispersion relation,
however it is not excluded by analyticity or unitarity.  Then the sum
rule is obtained by combining the dispersion relation with the low
energy theorem.

What could go wrong with this?  Absent any problems with
electrodynamics, the only weak point is ignoring the possible
polynomial in the dispersion relation.  Only the constant term
($c_{0}$) in the polynomial matters at $\nu=0$.  Usually limits on the
growth of amplitudes at high energies are invoked to exclude the
$\{c_{j}\}$ with $j\ge 1$.  However, they do not exclude the constant,
$c_{0}$.  If the integral in eq.~\ref{disp} diverged it would be {\it
necessary\/} to reformulate it by formally subtracting $f_{2}(0)$.  The resulting
integral would be more convergent, but now the constant $f_{2}(0)$
would appear in the relation.  However, even if the integral in
eq.~\ref{disp} converges, still the constant $c_{0}$ could be non-zero
and spoil the sum rule.  Such a constant is called, for historical
reasons, a ``$J=1$ fixed pole''.\cite{Fox:cc} So the question of the
validity of the DHGHY Sum Rule comes down to whether $J=1$ fixed poles
occur in QCD\null. It is known that they do not occur in low orders of
perturbation theory.  This was first verified when the electroweak
anomalous magnetic moment of the muon was calculated (for the first
time(!)) using a generalization of these
methods.\cite{Altarelli:1972nc}  Subsequently it has been studied to
higher orders.  Brodsky and Primack have argued that it does not occur
in ordinary bound states.\cite{Brodsky:1969ea} They show that the anomalous
magnetic moment of hydrogen can be calculated from a generalized
DHGHY Sum Rule with out a $J=1$ fixed pole.  Still, the verdict is
out in QCD, where bound states are not so simple.

If the DHGHY Sum Rule is verified experimentally, this question will recede to a footnote to history.  If, however, experiment fails to confirm it, we will all have a lot to learn about $J=1$ fixed poles!

\subsection{Orbital angular momentum in QCD?}

Even in the old days (pre-1988), it was clear that quark and gluon
spin distributions could be measured in deep inelastic scattering.  In
some uncertain sense they were imagined to be part of a relation which
gave the nucleon's helicity, $\fract12  = \fract12 \Delta\Sigma + \Delta g +
\hbox{[the rest]} $ where ``the rest'' was not well understood. $\Delta\Sigma$ and $\Delta g$ were (and are) measurable, gauge
invariant, and given by integrals over $x$ of well-defined quark and
gluon distribution functions.  Significant progress occurred in the
late '80s and '90s as the other pieces of the angular momentum were
related to local, gauge invariant operators.\cite{Jaffe:1990jz} This
line of work culminated in Ji's decomposition of the nucleon's
helicity,\cite{Ji:1997ek} $\fract12  =\fract12 \Delta\Sigma + \hat L_{q} +
\hat J_{g}$ where $\hat L_{q}$ is the nucleon matrix element of an
operator that rotates quarks' orbital motion about the $\hat
e_{3}$-axis in the rest frame.  $\hat J_{g}$ is the nucleon matrix
element of the operator that rotates the gluon about the $\hat e_{3}$
axis.  Ji showed that $\hat J_{g}$ cannot be further decomposed into
$\Delta g$ and an orbital contribution given by a local gauge
invariant operator.  This should not be too surprising because it is
well known that $\Delta g$ itself cannot be expressed in terms of a
{\it local\/} gauge invariant operator.\cite{Manohar:1991jx} [In
general the operator is non-local, but becomes local in $A^{+}=0$
gauge.]  The virtue of Ji's decomposition is that $\hat L_{q}$ can be
measured in deeply virtual Compton scattering (DVCS).  Although $\hat
J_{g}$ is in principle also measurable in DVCS, it requires a
precision study of $Q^{2}$ {\it evolution\/} DVCS data and is
impossible for practical purposes.

Most recently it has been possible to define gauge invariant {\it parton distributions\/} for all the components of the nucleon's angular momentum,\cite{Hagler:1998kg,Harindranath:1999ve,Bashinsky:1998if}
\be
	\fract12  =\int_{0}^{1}dx\left\{\fract12\Delta\Sigma(x,Q^{2})
   	+\Delta g(x,Q^{2}) + L_{q}(x,Q^{2}) +L_{g}(x,Q^{2})\right\}
	\la{newspin}
\ee
where $L_{q}$ and $L_{g}$ are Bjorken-$x$ distributions of quark and
gluon {\it orbital} angular momentum in the infinite momentum frame. $L_{q}$ and $L_{g}$ are given by the light-cone fourier transforms of
bilocal operator products just like other parton distributions.  This
decomposition has many virtues: the four terms evolve into one another
with $Q^{2}$,\cite{Hagler:1998kg,Harindranath:1999ve} each term is the
Noether charge associated with the appropriate transformation of
quarks or gluons.\cite{Bashinsky:1998if} Thus $L_{g}(x,Q^{2})$ is the
observable associated with the orbital rotation of gluons with
momentum fraction $x$, about the infinite momentum axis in an infinite
momentum frame.  On the other hand, eq.~(\ref{newspin}) suffers from a
significant drawback: unlike Ji's $\hat L_{q}$, we know of no way to
measure either $L_{q}(x,Q^{2})$ or $L_{g}(x,Q^{2})$.  They do not
appear in the description of DVCS.

So the situation with respect to a complete description of the
nucleon's angular momentum is frustrating.  The theory is under
control.  Eq.~\ref{newspin} summarizes all we would like to know, but
we do not know how to measure what we would like.  For a more complete
review, see the talk by X.~Ji in these proceedings.

\section{Noted in Passing\ldots}

\subsection{$g_{1}(x,Q^{2})$ at low-$x$} 
Reasonable extrapolations of existing data suggest that both $g_{1p}$
and $g_{1n}$ are negative and diverge as $x\to
0$.\cite{Altarelli:1998nb} This has little effect on the spin sum
rules, but belongs to the new, interesting low-$x$, strong coupling
regime of QCD\null.
	
\subsection{$s$ and $\bar s$ quarks in the nucleon}  
Strange quarks carry both momentum and spin in the nucleon.  The
strange quark's contribution to the nucleon spin can be extracted from
polarized DIS, hyperon $\beta$ decay data, and SU(3)
symmetry.\cite{Adeva:1998vw} The strange quark's contribution to the
nucleon's momentum is measured from dimuon production in neutrino DIS
\cite{Yu:1998su} These are both $C$-even observables and therefore add
quarks and antiquarks.\footnote{Parity violating ($xF_{3}$) dimuon
production in neutrino DIS can, in principle, separate $s$ and $\bar
s$ momentum distributions.  Present data are not accurate enough to
differentiate $s$ and $\bar s$.} Parity violating electron scattering
at low energy is sensitive to the strange quark contribution to the
nucleon's magnetization ($\mu_{s}$) and charge radius ($\langle
r^{2}_{s}\rangle$), both $C$-odd, and therefore sensitive to $s-\bar
s$.  Both SAMPLE\cite{Aniol:1999pn} and HAPPEX\cite{Spayde:2000qg}
report measurements, albeit with large uncertainties, consistent with
zero.  Theorists predict non-zero values for both $\mu_{s}$ and
$\langle r^{2}_{s}\rangle$ near the limits of experimental
sensitivity.  The next round of experiments should tell us whether $s$
and $\bar s$ quarks in the nucleon have significantly different
spatial distributions.

\subsection{Spin dependent quark gluon correlations in the nucleon}

The first high statistics measurements of the twist-three structure
function, $g_{2}(x,A^{2})$, for both the proton and neutron have been
reported by E155x at SLAC.\cite{Prepost:2000vt} For a complete review
see the talk of P.~Bosted in these proceedings.  The matrix elements
of quark-gluon correlations can be extracted from these
data.\cite{Shuryak:1981pi} They appear very small.  Perhaps dynamical
higher twist is always small.  If so, DIS data could be extrapolated
to very low $Q^{2}$ by including only kinematic higher twist.  This
approach, first suggested (for $g_{2}$) by Wandzura and
Wilczek,\cite{Wandzura:qf} would be a framework for connecting
low-$Q^{2}$ data with the DIS regime.  The generalized
Wandzura-Wilczek approximation offers the promise of a solid theoretical
foundations for speculations about ``parton-hadron duality'' in low
energy lepton scattering.
 
\subsection{Off-forward parton distributions}

Following the groundbreaking work by Ji,\cite{Ji:1996nm} the concept
of a parton distribution function has been generalized away from the
forward direction.  Ji, Radyushkin,\cite{Radyushkin:1997ki} and others
have shown that these ``skewed'' parton distributions can be measured
(with considerable effort) in deeply virtual Compton scattering
(DVCS).  I mentioned one physical application earlier in connection
with the concept of parton orbital angular momentum.  Many talks at
this conference are devoted to these new distributions.  What is
missing so far, to my taste, is a heuristic understanding of the
physical significance of off-forward parton distributions.  We still
need to figure out exactly what they are and what will we learn by
measuring them.

\subsection{Proliferation}

In response to new measurements of detailed properties in DIS
($k_{\perp}$ distributions, higher twist, semi-inclusive processes),
theorists have introduced a wonderful new zoo of distribution and
fragmentation functions.\cite{Mulders:2001jz} Now we need some clever
zoology to classify, relate, and interpret these new functions.  Are
they all independent or are they related to one another by as yet
unappreciated symmetries?  What is the general physical interpretation
of fragmentation functions analogous to the parton model of
distribution functions?  Perhaps some of them need to become extinct? How useful is the Wandzura-Wilczek approximation, which systematically
ignores dynamical higher twist?

\section{Conclusions}

My conclusions are brief.  
We have made striking progress in recent years.  The prospects for further
progress are excellent both in the immediate future and in the long term.

 \section*{Acknowledgements}

This work is supported in part by funds provided by the U.S.
Department of Energy (D.O.E.) under cooperative research agreement
\#DF-FC02-94ER40818 and in part by the RIKEN-BNL Research Center at
Brookhaven National Laboratory.

\end{document}